\documentclass[12pt]{article}
\usepackage{authblk}
\usepackage[utf8]{inputenc}
\usepackage[T1]{fontenc}
\usepackage[english]{babel}
\usepackage[sort]{cite}
\usepackage[a4paper,left=30mm,top=20mm,right=15mm,bottom=25mm]{geometry}
\usepackage{graphicx}
\usepackage{amsmath}
\usepackage{amssymb}
\usepackage{amsfonts}
\usepackage{latexsym}
\usepackage{setspace}
\usepackage{mathtext} 
\title{Giant self-induced transparency of intense few-cycle terahertz pulses in $n$-doped silicon}

\author[1]{\small O.V. Chefonov}
\author[1]{\small A.V. Ovchinnikov}
\author[1]{\small S.A. Romashevskiy}
\author[2]{\small X. Chai}
\author[2]{\small T. Ozaki}
\author[3]{\small A.B. Savel'ev}
\author[1]{\small M.B. Agranat}
\author[1]{\small V.E. Fortov}

\affil[1]{\footnotesize Joint Institute for High Temperatures of the Russian Academy of Sciences, Izhorskaya 13, Bd. 2, 125412 Moscow, Russian Federation}
\affil[2]{\footnotesize INRS-EMT, Advanced Laser Light Source, Varennes, Québec J3X 1S2, Canada}
\affil[3]{\footnotesize Lomonosov Moscow State University, Leninskie Gory, Moscow, 119991, Russian Federation}

\begin{document}

\date{\vspace{-5ex}}

\maketitle

\begin{abstract}
The results of high-field terahertz transmission experiments on $n$-doped silicon (carrier concentration of $8.7\times10^{16}$\,cm$^{-3}$) are presented. 
We use terahertz pulses with electric field strengths up to 3.1\,MV\,cm$^{-1}$ and a pulse duration of 700\,fs. 
Huge transmittance enhancement of $\sim$90 times is observed with increasing of the terahertz electric field strengths within the range of 1.5--3.1\,MV\,cm$^{-1}$.
\end{abstract}

Recent advances in intense THz science and technology have resulted in a number of subpicosecond THz sources with electric field strengths greater than 1\,MV\,cm$^{-1}$, which is now being exploited to directly excite transient states of matter in the nonlinear regime \cite{2}. 
Driving free carriers in a material far from equilibrium allows to study various scattering processes under extremely high electric fields on the picosecond timescales, without material breakdown \cite{3}. 
Distortion of potential profile by a strong applied field could result in tunneling processes between valence and conduction bands and along with impact ionization process induce carrier multiplication \cite{2,4}. 
In particular, hot carrier dynamics is of great importance to the semiconductor industry as the performance of ultrafast electronic devices relies on high-field transport phenomena. 
Among many semiconductors, silicon is widely used both in electronics and terahertz technology. 
Pure silicon is transparent in the far-infrared range with extremely low absorption below 0.05\,cm$^{-1}$ \cite{5}, while absorption of $n$-doped silicon is much higher ($n$-doped silicon is almost opaque for terahertz radiation at doping n$\sim$10$^{17}$ cm$^{-3}$).

Compared with other common semiconductors, there have been a few observations of nonlinear effects induced by intense THz pulses in doped silicon.
The first observations of the THz induced decrease in absorption by 10--20\% (or absorption bleaching) in $n$-doped bulk silicon have been demonstrated by nonlinear THz transmission and Z-scan measurements \cite{6,7}. 
The bleaching effect was attributed to the change in carrier mobility due to electron scattering into satellite valleys of the conduction band, which has higher effective masses and non-parabolic form.
The optical pump–THz probe measurements coupled to the intense THz source (up to 90\,kV/cm) on a silicon-based metamaterial have also shown the THz field-induced decrease in non-Drude-like conductivity due to intervalley scattering \cite{8}.

Recent experiments based on nonlinear THz transmission, nonlinear THz pump–THz probe and Z-scan measurements have allowed to monitor intervalley and intravalley dynamics of hot free carriers in bulk $n$-doped semiconductors such as gallium arsenide (GaAs), germanium (Ge) and silicon (Si), as well as the generation of new carriers by impact ionization in indium antimonide (InSb) at THz electric field strengths of up to 150\,kV\,cm$^{-1}$ in the 0.35--1.5\,THz range \cite{3,6,9}. 
Other THz-induced nonlinear effects have also been observed, including coherent emission at 2\,THz, nonlinear absorption bleaching and self-phase modulation in GaAs \cite{9,11,12}, saturable absorber behavior of GaAs, GaP, and Ge \cite{13}, coherent control of quantum states in InSb \cite{14}. 
However, the above-mentioned measurements have been carried out at electric fields below 1\,MV\,cm$^{-1}$ and changes in the THz transmission were ~20\% or lower.

In this paper, we studied transmission of $n$-doped silicon excited with intense subpicosecond terahertz pulses with electric field strength up to 3.1\,MV\,cm$^{-1}$. 
The sample used was a commercially available $n$-doped silicon wafer of 245\,$\mu$m thickness with carrier concentration and mobility of $(8.7\pm0.9)\times10^{16}$ cm$^{-3}$ and 800\,cm$^2$V$^{-1}$s$^{-1}$, respectively, obtained from Hall effect measurements.

\begin{figure}[ht!]
\centering
\includegraphics[width=0.6\linewidth]{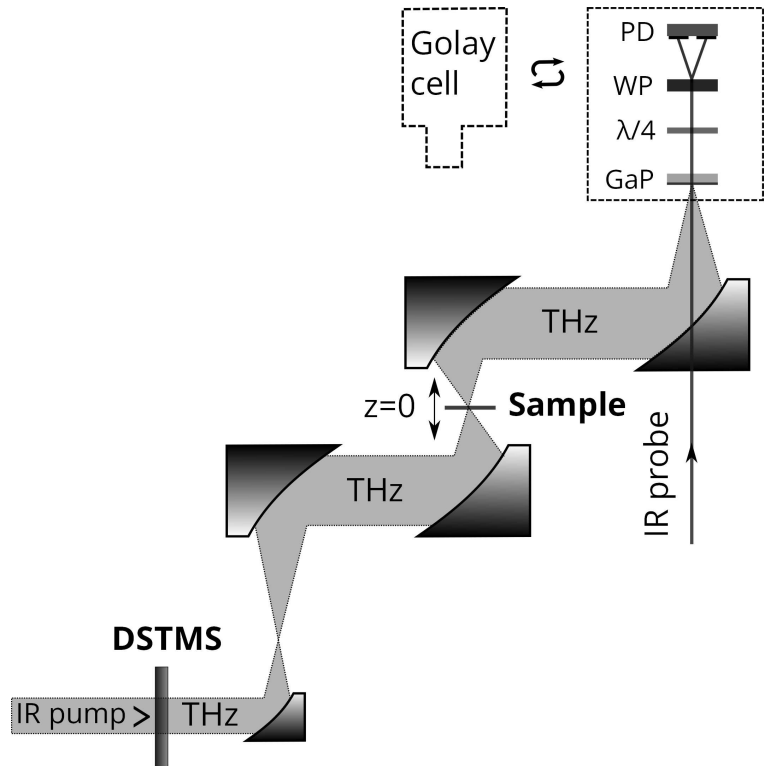}
\caption{Open-aperture Z-scan technique. PD --- balanced  InGaAs photodetector, WP --- Wollaston prism.}
\label{fig1}
\end{figure}

In this experiment, we used the open-aperture Z-scan technique, based on gradual varying of the terahertz radiation intensity on the surface of the sample moved along the beam line by a stepping motor, to study nonlinear effects \cite{8,9,15} (Fig.~\ref{fig1}).

Terahertz radiation was generated by the optical rectification in the nonlinear organic crystal DSTMS, pumped by of 100-fs femtosecond laser pulses at a central wavelength of 1240\,nm delivered by an amplified Cr:forsterite laser system \cite{16,17}. 
A broadband terahertz filter (LPF8.8-47, Tydex) with a cut-off wavelength of 34\,$\mu$m was placed after the crystal. 
Two off-axis parabolic mirrors were used as a 5:1 telescope to expand the terahertz beam. 
Finally, terahertz pulses with energy of 6.3\,$\mu$J were focused with an off-axis parabolic mirror with a 2\,'' diameter and a focus length of 2\,'' onto the sample to a spot with a radius of 200\,$\mu$m at $1/e^2$ measured using the knife edge technique. 
The duration $\tau_{FWHM}$ of the terahertz pulse measured with a first order autocorrelator (a THz Michelson interferometer) was 700\,fs at FWHM.

\begin{figure}[ht!]
\centering
\includegraphics[width=0.8\linewidth]{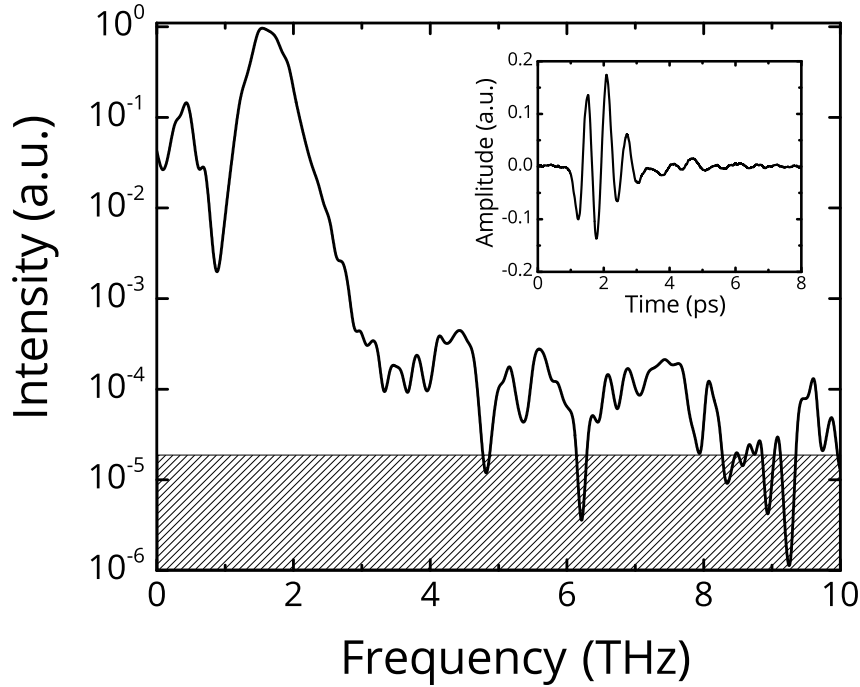}
\caption{The terahertz electric field in the time domain (inset) and its Fourier-transformed amplitude measured without the silicon sample. Noise level is shown as the shaded region.}
\label{fig2}
\end{figure}

The terahertz electric field after the silicon wafer was measured using electro-optical sampling in a nonlinear bilayer 2.1\,mm thick GaP crystal with a 2\,mm thick GaP~(100) and a 0.1\,mm thick GaP~(110). 
The time domain trace without the wafer and its Fourier‐transformed amplitude spectrum are shown in Fig.~\ref{fig2}. The spectrum extends over the range of 0.1--8\,THz with maximum at 1.6\,THz. 
The electro-optical measurements with the wafer at different positions along the terahertz beam waist did not show any considerable changes in the terahertz electric field and its spectrum. 

\begin{figure}[ht!]
\centering
\includegraphics[width=0.8\linewidth]{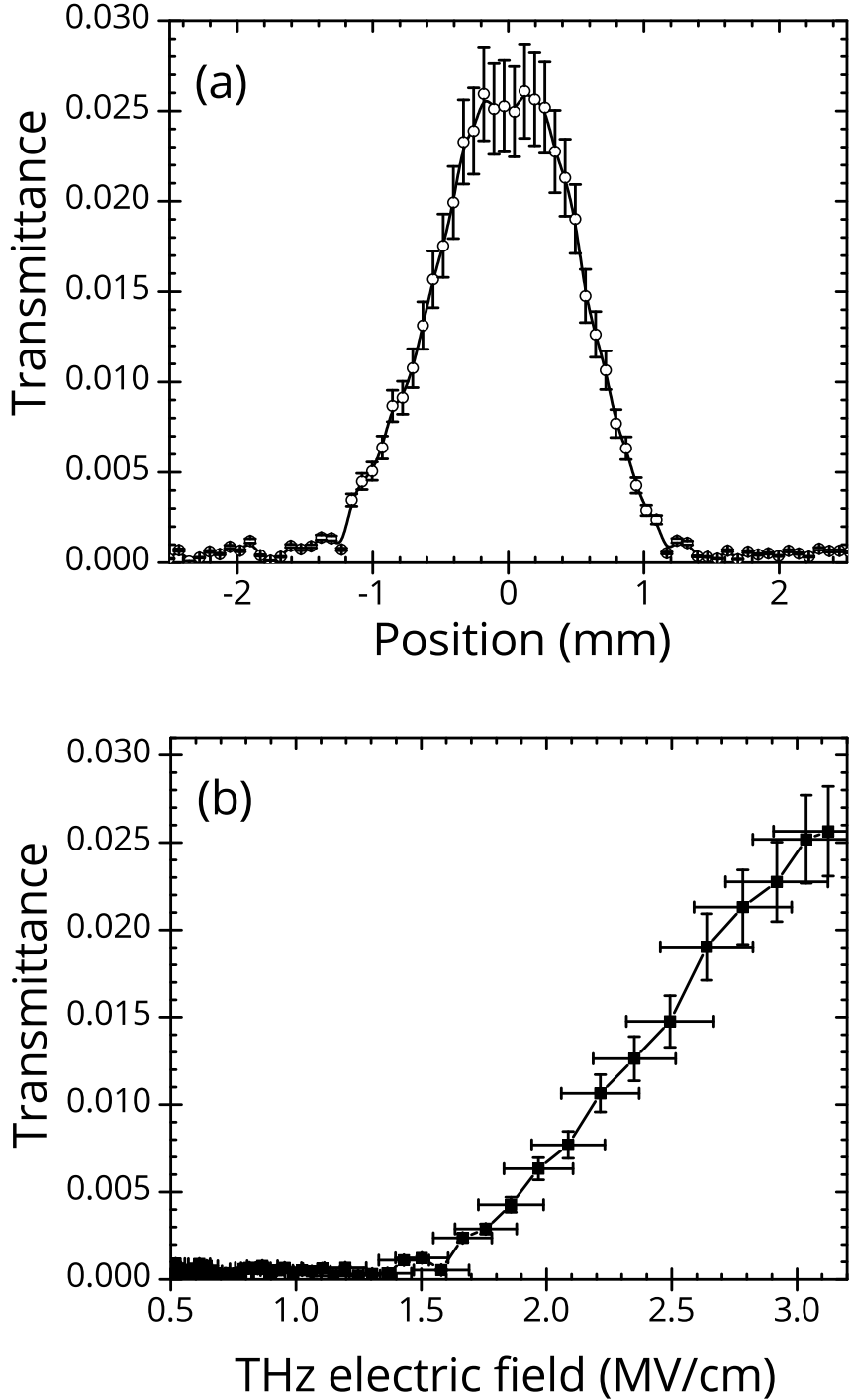}
\caption{Terahertz field-induced transmittance of $n$-doped silicon measured using Z-scan technique (a) and its deduced dependence on the terahertz electric field strength (b).}
\label{fig3}
\end{figure}

Transmittance of the terahertz radiation as a function of the silicon sample position in the Z-scan measurements is presented in Fig.~\ref{fig3}(a). 
Each point in the graphs was averaged over 10 measurements. 
The electric field strength of the terahertz pulse that corresponds to each position of the sample in Fig.~\ref{fig3}(a) were derived on the assumption that the terahertz radiation beam had the Gaussian energy transversal profile. 
In this case, the terahertz beam radius (at a level of $1/e^2$) over the distance Z is determined by:

\begin{equation}
w^{2}(Z)=w_{0}^{2}\left[1+\left(\frac{\lambda Z}{\pi w_{0}^{2}}\right)^2\right]
\end{equation}

and its intensity is:

\begin{equation}
I_{0}(Z)=\frac{4\sqrt{\ln2}}{\pi\sqrt{\pi}}\frac{W_{THz}}{w^{2}(Z)\tau_{FWHM}},
\end{equation}

where $w_{0}$ --- a minimal beam radius at the same level,  $\lambda$ --- central wavelength, $W_{THz}$ --- pulse energy. The electric field strength is then derived from:

\begin{equation}
E_{THz}(Z)\left[\frac{V}{cm}\right]=27.45\sqrt{I_{0}(Z)\left[\frac{W}{cm^2} \right]}.
\end{equation}

Fig.~\ref{fig3}(b) shows the dependence of the THz transmittance on the THz field strength $E_{THz}$. 

The transmittance was at the same level of 2$\sim$4$\times10^{-4}$ at $E_{THz}<1.5$\,MV\,cm$^{-1}$ and increased drastically by almost 90 times in the range of 1.5--3.1\,MV\,cm$^{-1}$ with linear slope of 0.02\,cm\,MV$^{-1}$. 
Note, that terahertz field-induced transparency in silicon with a carrier concentration of $\sim 10^{16}$\,cm$^{-3}$ was observed at much lower electric fields (16--135\,kV\,cm$^{-1}$), but transmittance increase was as small as 18\% \cite{7}. 
Saturated absorption observed in \cite{13}  also had a very low contrast, less than 30\%. 
In contrast, we observed threshold-like behavior of the transmittance of the $n$-doped Si, but at much higher electric fields.

In general, THz field induced scattering effects can affect significantly the nonlinear response of free charge carriers in materials \cite{18}. 
The dominant scattering mechanisms are determined by the properties of the material, the specific sample as well as the characteristics of the THz pulses \cite{18}. 
For example, intervalley scattering between non-energetic equivalent valleys governs the high field response of polar semiconductors \cite{19,20,21,22,23}. 
However, for covalent semiconductors such as Si, 6 equivalent valleys lie along [100] directions near the zone boundary \cite{24}. 
The intervalley scattering between equivalent valleys can thus play an important role on the macroscopic THz conductivities.

For the investigated Si sample, several scattering effects including impurity scattering, acoustic deformation and equivalent intervalley scattering could be the dominant mechanisms \cite{24}. 
At high fields, carriers can acquire relatively large kinetic energies and thus are less affected by the ionized impurities. 
Since several phonons are involved, intervalley scattering increases faster than the acoustic phonon scattering and dominate the scattering rate at high energies \cite{24}. 
Even though there is one rather high-lying ($L$) side valley that is accessible energetically (0.88\,eV above the band minimum), the effective conductivity mass of the $L$ valley is less than that of the $X$ valley. 
In addition to the relatively low mobility of electrons in the $X$ valleys, the overshoot effect is thus much less pronounced for Si compared with polar semiconductors such as GaAs.

As a result, we use a basic Drude plasma model to describe the THz complex dielectric function of doped Si and the simulated range of scattering time is chosen according to the intervalley scattering rate between equivalent valleys calculated by the Monte-Carlo method \cite{23,24}:

\begin{equation}
\varepsilon(\omega) = \left( n + \frac{i \alpha c}{2 \omega} \right)^{2}  = \varepsilon_{dc} - \omega_{p}^{2}/\left( \omega^{2} - i \omega/\tau \right).
\end{equation}

Here, $n$ and $\alpha$ are the frequency-dependent refractive index and absorption coefficient.   $\varepsilon_{dc}$ is the static dielectric constant of silicon and $\tau$ is the scattering time. 
$\omega_{p}= \left( Ne^2/{\varepsilon_0 m}\right)^{1/2}$  is the plasma frequency, where $N \approx 9\times 10^{16}$\,cm$^{-3}$ is the doping concentration and $\varepsilon_{0}$  is the vacuum permittivity. 
In our simulation, we decrease the total scattering time from 200\,fs down to 10\,fs for the investigated THz spectrum (0.1--6\,THz).

\begin{figure}[ht!]
\centering
\includegraphics[width=0.8\linewidth]{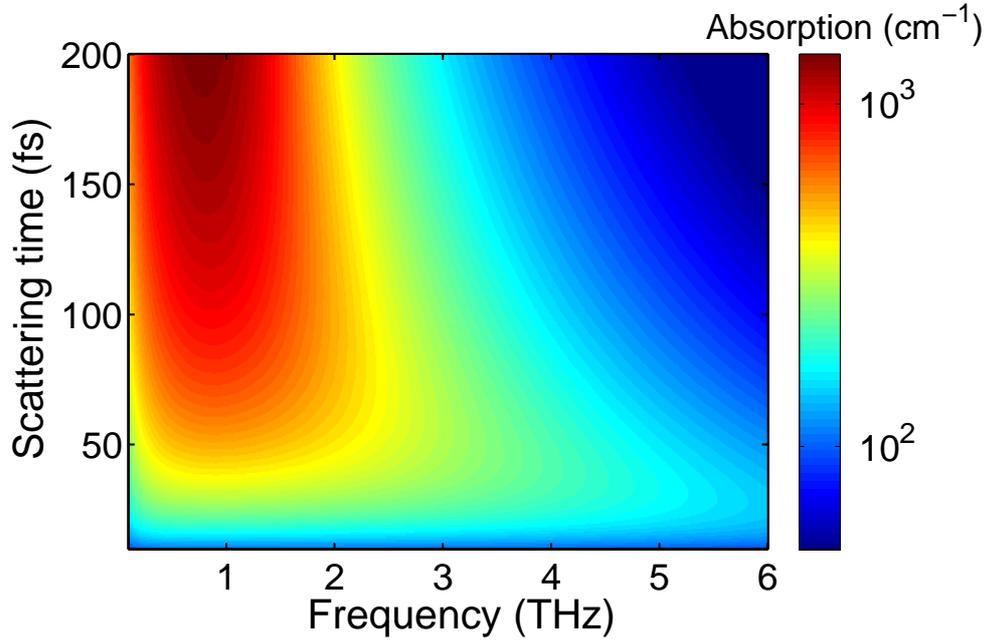}
\caption{Absorption coefficient of doped Si with decrease in total scattering time.}
\label{fig4}
\end{figure}

Intense THz transient can drive the carriers rapidly and in turn decrease the intervalley scattering time.  
As shown in Fig.~\ref{fig4}, the reduced total scattering decrease the macroscopic conductivity of Si, leading to a reduction of the power absorption coefficient over the studied spectrum. 
Nonlinear THz transmission enhancement is thus expected at high fields.

Hence, we demonstrated new experimental data on intense terahertz field-induced effects in $n$-doped silicon. 
The interaction of subpicosecond terahertz pulses at electric field strengths in the range of 1--3\,MV\,cm$^{-1}$ with doped silicon results in unprecedently high increase in its transmittance by almost two orders. 
This opens the path to THz saturable absorbers and new THz field controlled electronic devices. 
Quantitative description of the observed phenomenon demands more sophisticated modeling beyond the frame of the Dude model accounting for the real electronic structure of the $n$-doped Si and transient nature of the electron distribution function under action of a strong THz field.

\section*{Funding.}
Russian Science Foundation (RSF) (17-19-01261).

\section*{Acknowledgment.}
The experiments were performed using the unique scientific facility ``Terawatt Femtosecond Laser Complex'' of the Joint Institute for High Temperatures of the Russian
Academy of Sciences.

\end{document}